\documentclass[12pt,epsfig]{article}
\usepackage{amsmath}
\usepackage{graphicx}
\usepackage{epstopdf}
\usepackage{lscape}
\usepackage{longtable}
\usepackage[usenames, dvipsnames]{color}
\usepackage{caption}
\usepackage{subcaption} 
\usepackage{amssymb}
\begin{document}
	\begin{center}
		\textbf {Evaluation of the number of clusters in a data set using $p$-values from Multiple Tests of Hypotheses}\\
	\end{center}
	\begin{center}
		Dr. Soumita Modak$^{*}$\\
		Faculty,
		Department of Statistics\\ 
		University of Calcutta,
		Basanti Devi College\\
		147B, Rash Behari Ave, Kolkata- 700029, India\\
		Email: soumitamodak2013@gmail.com\\
		Orcid id: 0000-0002-4919-143X\\
		Homepage: sites.google.com/view/soumitamodak
		
	\end{center}
	\begin{abstract}
		This paper proposes a novel, nonparametric, interpoint distance-based measure to investigate whether there exist any groups in a set of given data, and if so then, how many groups are prevailing in total. It is a cluster accuracy index useful for arbitrary-dimensional data set, in association with any clustering algorithm having the number of groups specified as a priori. We perform univariate, nonparametric, multiple statistical tests of hypotheses, where as many dependent tests as the sample size are carried out using the interpoint distances. They possess $p$-values to be combined to reach a decision, which is taken in a step-wise process for a possible number of clusters. It reduces the unnecessary computations compared with the other accuracy measures from the literature. Data study establishes the proposed index's efficiency and superiority.
	\end{abstract}
	keywords: {Cluster accuracy index; Investigation into grouping; Evaluation of number of clusters; Nonparametric multiple tests of hypotheses; Interpoint distance; High-dimensional applicability.}
	\section{Introduction}
	Cluster analysis, one of the most widely used statistical data mining techniques, finds the inherent groups in a given set of data and helps explore the characteristics regarding them (Ester et al. 1996; Jain et al. 1999; McLachlan and Peel 2000; Everitt, Landau and Leese 2001; Campello, Moulavi and Sander 2013; Modak 2019; Modak 2021; Modak 2023a, 2024a, 2024b). One of the common challenges in clustering algorithms, like $K$-means (Hartigan 1975; Hartigan and Wong 1979; Modak, Chattopadhyay \& Chattopadhyay 2018), $K$-medoids (Kaufman and Rousseeuw 2005; Modak, Chattopadhyay \& Chattopadhyay 2017, 2020), hierarchical (Kaufman and Rousseeuw 2005; Johnson and Wichern 2007; Modak, Chattopadhyay \& Chattopadhyay 2022), parametric mixture-model based (McLachlan and Peel 2000; Tarnopolski 2019; Toth et al. 2019), is to provide the unknown value of the present clusters (denoted by $K$) as a priori. For this purpose, a lot of efficient cluster validity measures are available like Cali\'{n}ski and Harabasz index (Cali\'{n}ski \&  Harabasz 1974), Dunn index (Dunn 1974), average silhouette width (Rousseeuw 1987), nearest neighbor classification error rate (Ripley 1996), connectivity (Handl et al. 2005), indices proposed by Modak (Modak 2022, 2023b, 2023c), and many others (see, for example, Ball and Hall 1965; Davies and Bouldin 1979; Fraley \& Raftery 1998; Pakhiraa, Bandyopadhyay and Maulik 2004; Cheng et al. 2019; Silva et al. 2020; and references therein). However, most of the existing measures work with the assumption that the true value of $K$ is greater than 1, and hence do not check for the possibility that $K$ may be just equal to 1. Therefore, when actually $K=1$, such approaches can land in misleading clustering or just binary classification, indicated by the optimal values of the corresponding indices, as a consequence of the induced partitions with $K>1$.
	
	Measures that take care of the above include the classic Bayes'
	information criterion (BIC), Bayes factor (Schwarz 1978; Kass and Raftery 1995; Frayley and Raftery 1998; T\'{o}th, R\'{a}cz \& Horv\'{a}th 2019), gap statistic (Tibshirani, Walther, \& Hastie 2001; Modak, Chattopadhyay \& Chattopadhyay 2018). However, none of these are distribution-free. BIC and Bayes factor are inherently designed involving the parametric mixture models under consideration. Tibshirani, Walther, \& Hastie (2001) adopt a
	computationally-extensive reference distribution for determination of $K$, which
	decides the sampling distribution of their gap statistic. They select a data-driven reference distribution through bootstrapping. There a large number of samples is simulated from a uniform distribution
	on the hypercube, determined by the ranges of data centered and then transformed by the linear principal component analysis (Sch\"{o}lkopf \& Smola 2002; Modak, Chattopadhyay \& Chattopadhyay 2018). Thus,
	it is computationally quite time consuming and depends on the
	parametric model assumptions regarding the reference distribution. Moreover, in spite of using the interpoint distances, the computation technique makes its application questionable to the high-dimensional situations. Likewise, as the implementation of BIC or Bayes factor requires the supposition about a parametric model and subsequently the estimation of the concerned parameters, the high-dimensional data pose a real challenge (see, for details, Bai
	\& Saranadasa 1996; Ahn et al. 2007; Jung \&
	Marron 2008; Yata \& Aoshima 2010; Marozzi 2015). On the contrary, our measure is purely based on the interpoint distances without any model or parametric assumptions, and hence applicable, with equal expectation of performance, to any dimensional space. This index implements the nonparametric, dependent, multiple tests of hypotheses for each of the observations. The tests are applied to the interpoint distances, computed for an observation, from the cluster it is assigned to, and from its nearest cluster (as defined by the analyst). Then we combine all the resulted $p$-values, obtained from the multiple tests, to reach the final decision on the value of $K$ in a step-wise procedure. We consider all the possible values for $K=1,2,...,$ etc., one by one, and reach an appropriate value or stop at a suitable step, that saves us much computations compared to the other accuracy indices. 
	
	The present paper is constructed in the following way. Section~2 contains our proposed method and discusses its various aspects. In Section~3, we perform our data study and show our measure's efficacy in comparison to its competitors, and Section~4 draws the conclusion. 
	\section{Proposed measure}
	Suppose we are given a data set having $N$ members or observations in any-dimensional real space, where the proximity between any 2 data members is measured in terms of some suitably chosen distance, popularly referred to as `interpoint distance' (Jure\v{c}kov\'{a} and Kalina 2012; Marrozi 2016; Modak 2022).
	\subsection{Objective}
	For a given set of data, our goal is to first check whether there exists any grouping intrinsic to the data set, and if it does then, how many groups, mutually exclusive and exhaustive, are prevailing. We propose a novel interpoint-distance based cluster validity measure, which can assess the output generated from any unsupervised classification algorithm, where $K$ is provided as a priori, and thus serves our objective.
	\subsection{Intuition}  
	The concept lies in the fact that if there is no clustering pattern present in the data set, then partition of the data would not result in any separate groups other than mere divisions of the same data set without any meaningful significance. Therefore, we apply a test to check the hypothesis that the populations of 2 groups, achieved through clustering with $K=2$, are the same or not. Such a test is carried out under a nonparametric set-up for each member of the data in terms of its interpoint distances, from the cluster that particular member is clustered to (represents our population 1) and from the other cluster, i.e. its nearest cluster as defined by the analyst (considered as the second population). Finally we combine the results of all such tests for reaching a decision on whether the data set under analysis has only 1 cluster as a whole when we expect the null hypotheses to be true, or the alternatives are true which indicate clustering of given data with at least 2 clusters existing. Now, using our combined resolution from the multiple tests of hypotheses, either we conclude with 1 group, or at least 2 classes are established. In the latter case, we further go on testing whether the found 2 groups are the only kinds present or they can be further split into at least 3 clusters, adopting the same logic. In this stage, we have data clustered into 3 classes (i.e. $K=3$ is provided), and we check the equality of 2 populations corresponding to each of the member's mother cluster and its determined nearest cluster in terms of the computed interpoint distances. Thus, at the present step, acceptance of the null hypotheses indicate that the additional cluster (with $K=3$) is unnecessary, and we conclude with 2 groups. Otherwise, the mentioned process is repeated step-wise for $K=4,5,6,...$, etc., until a final grouping of data is reached.
	\subsection{Methodology}
	The above-mentioned multiple tests corresponding to each of the $N$ observations from the data are implemented in the form of the well-known homogeneity tests, from the nonparametric domain, under a univariate set-up, with a specified value of $K=2,3,...,$ etc. The test statistic for testing the hypothesis of equality of several proportions for the populations under study (here 2 populations), known as the chi-square test statistic for test of homogeneity (Hogg, Mckean and Craig 2019), follows an approximate chi-square distribution, with degree of freedom equal to (the number of similarly defined categories in both the populations) minus 1. Here the interpoint distances from the 2 concerned clusters are normalized and then the union of their ranges $(\in (0,1])$ is divided into a number of equally-spaced, mutually exclusive and exhaustive intervals, which represent the different categories in the populations duo. The number of categories is denoted by ($w-1)$, where the minimum value of $w$ is 3 that corresponds to a $2\times2$ contingency table and leads to the least possible computation. We should vary the value of $w$ and observe its impact on our measure. For a fixed value of $K=2,3,...$, etc., we consider the $p$-values obtained from the $N$ number of dependent multiple tests of hypotheses corresponding to each of the members. The literature has many suggestions for combining such $p$-values, however, we select the simple one by giving all the tests equal weights and taking the simple arithmetic mean of the $p$-values to compute our integrated $p$-value. For $K=K'(>1)$ given as a priori in our step-wise procedure, the true value of $K$ as $(K'-1)$ is accepted for an
	integrated $p$-value less than the nominal level of significance $\alpha$.
	\subsection{Algorithm}
	(a) Cluster the data set of size $N$ using a clustering algorithm with $K=2$ provided.\\
	(a1) Consider the member $i$ from the clustered data set.\\
	(a2) Compute the interpoint distances from $i$ to all other members of the data set.\\
	(a3) Find the nearest cluster of $i$. We follow the widely used concept for defining that (Rousseeuw 1987; Kaufman and Rousseeuw 2005; Modak 2022) as the one, among all non-mother clusters, having a minimum average of distances, computed from $i$ to all members of that cluster. However, for $K=2$, the choice of nearest cluster is obvious.\\ 
	(a4) Therefore, one population is represented by the sample of interpoint distances from $i$ to all other members of its mother cluster, and the other by distances between $i$ and all members of its nearest cluster, under the homogeneity test. For this, the interpoint distances from the 2 concerned clusters are normalized and then the union of their ranges $(\in (0,1])$ is divided into a number of equally-spaced, mutually exclusive, and exhaustive intervals. These intervals represent the similarly formed categories in the populations duo for which we count the frequencies independently. Here the normalization and the range selection for the distances are made as an attempt to avoid having too small frequencies in the categories that can hinder our test statistic from attaining its chi-square distribution. Now we implement the test of homogeneity between the mentioned populations and record the corresponding $p$-value as $p_i$.\\
	(a5) Repeat steps (a1-a4) for each member $i=1,...,N$, performing $N$ number of multiple tests for homogeneity and record the dependent $p$-values $p_i$ for $i=1,...,N$.\\ 
	(a6) Final conclusion, regarding given data set has 1 or at least 2 clusters, is taken by combining the decisions from the tests for all the members, in terms of an integrated $p$-value, denoted by IP. It is based on a $p$-value $p^{*}$ that possesses,
	\begin{equation}
		p^*=p_i\hspace{.05in}\text{with}\hspace{.05in}\text{probability}\hspace{.05in}1/N,\hspace{.05in}\text{for}\hspace{.05in} i=1,...,N,
	\end{equation} 
	under the multiple null hypotheses and the alternatives. This leads to the following:
	\begin{equation}
		\text{IP}=E(p^*)=\sum_{i=1}^{N}p_i/N, 
	\end{equation}
	where 
	\begin{equation}
		Prob\hspace{.05in}[p^*<\alpha]=\sum_{i=1}^{N}Prob\hspace{.05in}[p_i<\alpha]/N.
	\end{equation}
	Consequently, the algorithm terminates with the resolution of a single cluster (i.e. $K=1$) for a value of IP $\geq \alpha$, otherwise we go to the next step.\\
	(b) Cluster the data set of size $N$ using a clustering algorithm with $K=3$ provided.\\
	(b1) Analogously, repeat all the steps (a1-a6) to make the verdict on the true number of clusters existing is $2$ (the null hypotheses) or at least $3$ (the alternatives).\\
	(b2) Similarly, the algorithm ends with existence of 2 groups for a value of IP $\geq \alpha$, otherwise we go to the subsequent step.\\
	(c) Likewise we proceed step-by-step with clustered data for larger values of $K=4,5,...,$ etc., as a priori, until $K=K'$ (say), while the null hypotheses are accepted. Then, the algorithm stops with the true value of $K$ estimated as $(K'-1)$.
	
	Now, we discuss the computation of our integrated $p$-value.
	As described above, for the given data clustered into a specified number of clusters, we perform multiple statistical tests and obtain the corresponding $p$-values as many as our data size. Individuals of these tests of hypotheses may lead to different conclusions; however, the final decision should be taken by combining these dependent $p$-values appropriately. Various methods are available in the literature to combine such independent or dependent $p$-values (see, for example, Kost and McDermott 2002; Poole et al. 2016; Modak and Bandyopadhyay 2019; and references therein). However, there exist constraints in the forms of the preference for acceptance of a particular decision, the multivariate dependence structure of the tests under analysis and so on. Therefore, we consider an integrated $p$-value in terms of the simple arithmetic mean computed over all individual $p$-values, where we assign equal importance to all of the tests by weighing them equally (as explained in step (a6) above).
	
	It is known that the homogeneity test adopts the chi-square approximation regarding the distribution of the test statistic. However, it may not stand valid if the frequency of any category is found to be small enough, usually less than 5. Therefore, for large values of $w$ and $K$, to avoid such cases, we implement a Monte Carlo simulation technique with `$B$' replicates (for details, see, Hope 1968) to compute the $p$-value for each of the multiple tests of hypotheses. In our work, $B =10,000$ is maintained throughout to achieve robust estimates.
	\subsection{Properties}
	(i) Our method can determine $K$ in a data set for any of the original value for $K$ greater than or equal to 1, in contrast to many other popular indices like Dunn index, average silhouette width, connectivity, etc., which do the task only with the assumption that $K>1$.\\ 
	(ii) Our approach evaluates the true (unknown) value of $K$ for a data set in a step-wise process, where at each step we use the data clustered into $K$ (a priori) clusters, with $K$ increased serially starting from 2. Therefore, any clustering algorithm, with specified hyperparameter $K$, is compatible with our measure.\\
	(iii) One major advantage of our measure is it does not require to be computed for a preassigned set of values for $K$, unlike most of the other indices like BIC, gap statistic, Dunn index, average silhouette width, where the true value of $K$ is estimated as the one with an optimal value for a measure computed over a fixed set of values for $K$; whereas our index is only computed for the clustered data till a value of $K=K'$ at which our designed algorithm ends with estimated $K$ as $(K'-1)$. Thus we can significantly reduce the computational burden while implementing our index. Also the selection of the highest value of $K$ till which the measures are to be computed, i.e. the highest possible number of clusters in a data set which is actually unknown in a real-life situation and difficult to presume correctly for the analyst, is avoided by our automated step-wise method.\\
	(iv) We use a concept of nearest cluster in the analysis, where we follow the widely used definition of the nearest cluster of a member $i$ in the clustered data as the one among all its non-mother clusters having a minimum average of distances, computed from $i$ to all members of that cluster. Here, any other notion about the nearest cluster could also be used; for instance, we apply median of distances in the place of average distance, wherein data study (see, the study of (S3) in the presence of outliers, under Section 3) exhibits the robustness of our measure with respect to the formation of nearest cluster.\\
	(v) Our nonparametric, distribution-free, validity index is based on interpoint distances. Hence it can be applied to any kind of data by the help of choosing a proper distance measure, whereas the utilization of the distance makes our index efficiently applicable to any-dimensional space.
	\section{Data study}
	We study synthetic data sets under 3 distinct settings, denoted by S1-S3. As the results can vary from one random sample to the other, we carry out 100 replications of simulation under each of the set-up. The efficacy of every measure is quantified in terms of the percentage of times, correctly estimating the true $K$, out of all the replications performed. Here, throughout our simulation study, we consider $\alpha=0.01$.
	
	In each case, we examine 3 competitors, namely the BICs based on 2 different models and the gap statistic, to compare the performance of our proposed index. With BIC, we employ 2 model-based clustering algorithms using parameterized finite Gaussian-mixture models as: (a) `EII': spherical with equal volume, and (b)
	`EEE': ellipsoidal with equal volume, shape and orientation (Celeux \& Govaert 1995; Fraley \& Raftery 1999; Scrucca et al. 2016). Covariance matrices respectively involve 1 parameter and [dimension$\times$(dimension+1)$\times (1/2)$] number of parameters for distribution from each of the components in the mixture models EII and EEE (Fraley and Raftery 2007). Model parameters are estimated by the maximum likelihood estimators through the expectation–maximization algorithm (Dempster et al. 1977; McLachlan \& Peel 2000), where the initialization is implemented with the hierarchical model-based agglomerative clustering (Banfield and Raftery 1993; Scrucca et al. 2016), for different considered values of $K\geq1$. In clustering both under (a) and (b), the best possible
	model is retrieved by a maximum value of the computed BIC (Fraley and Raftery 2002; Scrucca et al. 2016; and references therein). Another competitor is the popular gap statistic (Tibshirani, Walther, \& Hastie 2001; Modak, Chattopadhyay \& Chattopadhyay 2018), whose sampling distribution is derived from a reference distribution using bootstrapping (Efron and Tibshirani 1993). A large number of samples is simulated from a uniform distribution
	on the hypercube, determined by the ranges of the given data, which are first centered and then transformed by the linear principal component analysis (Sch\"{o}lkopf \& Smola 2002; Modak, Chattopadhyay \& Chattopadhyay 2018). Here the statistic is computed for different values of $K\geq1$, using the clustering output from any algorithm with $K$ as a priori. Its highest value suggests the optimal classification. For clustering, with our index and gap statistic, we conduct $K$-means method using `Hartigan-Wong' algorithm (Hartigan 1975; Hartigan and Wong 1979; Modak, Chattopadhyay \& Chattopadhyay 2018) and $K$-medoids method with `PAM' algorithm (Kaufman and Rousseeuw 2005; Modak, Chattopadhyay \& Chattopadhyay 2017, 2020).
	
	(S1) Firstly, we consider a multivariate normal population in a 5-dimensional space, with the expectation null vector and the correlation matrix having all off-diagonal entries as 0.5. Here, 2 different cases, one with a single cluster and the other with binary clusters, are analyzed.\\ 
	(S1a) In the former case, under each replication, we draw a random sample of size 250.\\
	(S1b) The latter case possesses 2 natural clusters, where the second cluster is created by adding $(0,1,-1,3,-3)'$ to half (i.e. 125) of the samples from (S1a) in every replication.
	
	Our index and its rival gap statistic are implemented with the widely-used $K$-means method, which happens to be well-performing for exposing such Gaussian clusters. Results are reported in Table~\ref{t1}. As expected, since the clusters from the finite Gaussian-mixture models are known to be ellipsoidal (Scrucca et al. 2016), BIC based on the model EEE comes out to be the best. On the other hand, BIC along with the model EII completely disappoints. While our novel measure is absolutely competitive, with consistency for different values of $w$ (denoted by IP$_w$), the gap statistic shows mediocrity. 
	
	Next, we consider some high-dimensional cases under set-ups (S1a$'$) and (S1b$'$), which are analogous to the above-described (S1a) and (S1b) respectively, with only difference in data dimension as 300, which is greater than the cluster size(s). The second cluster in (S1b$'$) is constructed by inserting a difference to half of the samples from (S1$a'$) obtained under each replication, where $(0,1,-1,3,-3)'$ is added to the first 5 dimensions of the sample values, then $(0,1,-1,3,-3)'$ is added to the next 5 dimensions (i.e. from the 6-th to the 10-th dimensions), and so on. Our index shows quite covetous outcome (see, Table~\ref{t1}); whereas its competitors lose their efficacy in high-dimension (for detailed discussions, see, Bai
	\& Saranadasa 1996; Fraley \& Raftery 2003; Ahn et al. 2007; Jung \&
	Marron 2008; Yata \& Aoshima 2010; Marozzi 2015). For computing the model-based BICs, we use the `mclust' package, version 5, in `R' programming language (Scrucca et al. 2016); where BIC with model EEE is not available for dimension equal to or greater than the sample size, and hence its value is not reported and marked as NA (Fraley et al. 2012; Scrucca et al. 2016).
	
	As for $w=3$, IP has the minimum possible computation (corresponds to a $2\times2$ contingency table) and the highest possible accuracy (due to implementing the approximate chi-square distribution), and possesses our desired results (experienced by the last data study), we carry out the next simulations for IP$_{w=3}$.\\
	(S2) Here the samples own 3 groups, each of size 200, from a quadrivariate non-normal population (Vale \& Maurelli 1983), specified through the correlation matrix having off-diagonal elements 0.6, skewness ($b_1$) = 1.75, and kurtosis ($b_2$) = 3.75 (Joanes \& Gill 1998). 
	
	Table~\ref{t2} reports the performances of IP, for $w=3$, and of the gap statistic, in association with 2 independent clustering algorithms, namely $K$-means and $K$-medoids, and of the model-based BICs. The results clearly depict that the existing clusters in the present data set are extremely challenging to be exposed. However, our index, with both clustering methods, outperforms the others.
	
	(S3) Lastly, we take into account the effects of outliers or extreme observations on our validity index. For this purpose, we consider the leptokurtic exponential distribution with the rate = 1, and study the cluster analysis of samples, drawn from this distribution with binary groups of equal sizes 250, in a bivariate space. Here we implement the $K$-means algorithm and the robust-against-outliers method $K$-medoids. Both are used with a tweak in defining the nearest cluster using the minimum median, along with the original mean version (see, the property (iv) under Section 2.5), as medians are less affected by outliers than means.
	
	Our measure IP (with $w=3$) is competitive enough with the gap statistic that performs with full efficacy, whereas the others entirely fail (see, Table~\ref{t3}). Moreover, the close enough conclusions by both the algorithms, and the exact same output from differently defined nearest clusters (i.e. IP$_w$(mean) and IP$_w$(median) for mean and median versions respectively), manifest the consistency of results obtained by our index.
	
	Now, we apply our technique to real observed data sets, say D1 and D2, as follows:\\
	(D1) We consider the Swiss banknotes data that contain observations on 6 variables for the old-Swiss 1000-franc bank notes with 2 groups of 100 genuine and 100 counterfeit (Flury and Riedwyl 1988). The study variables are: (i) length of bill, (ii) width of left edge, (iii) width of right edge, (iv) bottom margin width, (v) top margin width, and (vi) length of diagonal, where all are measured in mm.
	
	Our measure IP$_{w=3}$ (mean, median), for both values of $\alpha=0.01$ and $0.05$, with $K$-means and $K$-medoids methods, estimate $K$ as 2, which is the true value of $K$ here. We report the computed IPs, for clustered data with $K=2,3$ as a priori respectively, for $K$-means as: (0.00367, 0.14156) and for $K$-medoids equal to: (0.00523, 0.05726). Clearly, for $K=2$, IPs are less than $\alpha$, and for $K=3$, they exceed $\alpha$, which imply estimated $K$ is 2. 
	
	Here the adjusted Rand index ($\in[-1,1]$, see, Hubert and Arabie 1985) is also used to assess the clustering accuracy by measuring the agreement between the clustered data and the data with original classes. A higher positive value of it indicates better agreement. In here, the produced values are 1 and 0.96020 respectively, for the above 2 algorithms with $K=2$ specified. It indicates $K$-means method is better suited for clustering this data set than $K$-medoids. Therefore, we further report the results from our measure for $K$-means, with varying $w$. Consistently, IP$_w$(mean,median), for both values of $\alpha$, reach the correct decision on revealing the true value of $K$ as 2 as follows: (a) $w=4$, IPs: (0.00028, 0.0764), (b) $w=5$, IPs: (0.00021, 0.05944), (c) $w=6$, lacks sufficient category-frequencies.
	
	On the other hand, the gap statistic with $K$-means and $K$-medoids clusterings, and the BICs based on models EII and EEE respectively give estimated $K$ = (6, 5, 6, 3). That is, all the competitors fail in clustering this data set properly.\\
	(D2) In astrophysics, the 2-phase formation theory of nearby massive early-type galaxies (abbreviated to ETGs) is challenged by 3 possible different sources corresponding to the 3 components of nearby ETGs, namely innermost, intermediate and outermost (De et al. 2014). Thus, the so far established model stating the same kind of origins behind the latter duo components is disturbed by these findings. However, later on, Modak, Chattopadhyay \& Chattopadhyay (2017) solve the conflict using spatial data. Here, we apply our measure to investigate whether there are any subgroups in the combined data of intermediate and outer components of the ETGs. We perform the analysis in a bivariate space of mass and radius as our study variables (for details, see, De et al. 2014; Modak, Chattopadhyay \& Chattopadhyay 2017; and references therein).
	
	IP$_{w=3}$ (mean, median) estimate $K=1$ through $K$-means and $K$-medoids algorithms. As the corresponding IPs, computed using the clustered data with $K=2$ as a priori, are: 0.07339 and 0.07565 (greater than $\alpha=0.01,0.05$) respectively. The gap statistic with these clustering methods and the BIC with EEE also indicate estimated $K$ as 1, whereas the BIC based on EII wrongly suggests 3 clusters. Thus the well-known 2-phase formation of nearby ETGs is reconfirmed by our study.                    
	\section{Conclusion}
	We propose a novel nonparametric cluster accuracy measure to assess whether there exist any separate groups in a set of data given and if so how many groups there are in total, with the help of implementing univariate nonparametric multiple statistical tests of hypotheses applied to observation-wise interpoint distances. As many dependent tests as the number of observations are carried out whose $p$-values are integrated to take a decision in a step-wise process for each considered value of the possible number of clusters which is increased one by one as required. It reduces the unnecessary computations compared with the existing measures which have to be quantified simultaneously for a specified set of values for $K$. Data study establishes the proposed index's efficiency, superiority and high-dimensional applicability, whereas its wide usability is manifested by implementation with any clustering algorithm where the unknown number of groups is provided as a priori. The interpoint distances make it compatible with data of arbitrary dimension and having measurements on any scales with the help of selection of an appropriate distance measure, which can be studied in detail.
	\clearpage
	\begin{table}
		\caption{Clustering performance of different cluster assessment indices under the simulations in (S1)}
		\begin{center}
			\begin{tabular}{c|c|c|c}
				\hline
				Set-up&Cluster&Measure&Efficacy(\%)\\
				&algorithm&&\\\hline
				(S1a)&$K$-means&IP$_{w=3}$&100\\
				&,,&IP$_{w=4}$&99\\
				&,,&IP$_{w=5}$&100\\
				&,,&Gap&84\\
				&EII&BIC&0\\
				&EEE&BIC&100\\\hline

				(S1b)&$K$-means&IP$_{w=3}$&97\\
				&,,&IP$_{w=4}$&99\\
				&,,&IP$_{w=5}$&*\\
				&,,&Gap&43\\
				&EII&BIC&0\\
				&EEE&BIC&100\\\hline
				(S1a$'$)&$K$-means&IP$_{w=3}$&100\\
				&,,&IP$_{w=4}$&96\\
				&,,&IP$_{w=5}$&*\\
				&,,&Gap&68\\
				&EII&BIC&0\\
				&EEE&BIC&NA\\\hline
				
				(S1b$'$)&$K$-means&IP$_{w=3}$&100\\	
				&,,&IP$_{w=4}$&98\\
				&,,&IP$_{w=5}$&*\\
				&,,&Gap&0\\
				&EII&BIC&0\\
				&EEE&BIC&NA\\\hline
				
				\hline	\hline
				(*) many replications lack\\ enough category-frequencies\\NA: not available\\
			\end{tabular}
		\end{center}
		\label{t1}
	\end{table}
	\clearpage
	
	\begin{table}
		\caption{Clustering performance of different cluster assessment indices under the simulations in (S2)}
		\begin{center}
			\begin{tabular}{c|c|c}
				\hline
				Cluster&Measure&Efficacy(\%)\\
				algorithm&&\\\hline
				$K$-means&IP$_{w=3}$&	30\\
				$K$-medoids&IP$_{w=3}$&	28\\
				$K$-means&Gap&3\\
				$K$-medoids&Gap&3\\
				EII&BIC&0\\
				EEE&BIC&0\\
				\hline
			\end{tabular}
		\end{center}
		\label{t2}
	\end{table}
	\clearpage
	\begin{table}
		\caption{Clustering performance of different cluster assessment indices under the simulations in (S3)}
		\begin{center}
			\begin{tabular}{c|c|c}
				\hline
				Cluster&Measure&Efficacy(\%)\\
				algorithm&&\\\hline
				$K$-means&IP$_{w=3}$(mean)& 89\\
				$K$-means&IP$_{w=3}$(median)&89	\\
				$K$-medoids&IP$_{w=3}$(mean)&91	\\
				$K$-medoids&IP$_{w=3}$(median)&91	\\
				$K$-means&Gap&100\\
				$K$-medoids&Gap&100\\
				EII&BIC&0\\
				EEE&BIC&0\\
				\hline
			\end{tabular}
		\end{center}
		\label{t3}
	\end{table}
	\clearpage
	\section*{Disclosure statement}
	No potential conflict of interest was reported by the author.
	\section*{Acknowledgments}The author expresses her sincere gratitude to the Editor-in-Chief and thank the anonymous reviewer for providing expert feedback on the work which led to superior presentation in the revised form.
	

\clearpage
	\begin{thebibliography}{}
		\bibitem{}
		Ahn, J., Marron, J. S., Muller, K. M., Chi, Y.-Y. (2007). \textsl{The high-dimension, low-sample-size geometric representation holds under mild conditions}, Biometrika, \textbf{94},  760--766.
		\bibitem{}
		Bai Z. and Saranadasa H. (1996). \textsl{Effect of high dimension: by an
			example of a two sample problem.} Stat Sinica, \textbf{6}, 311–-329.
		\bibitem{}
		Ball,  G. H. and Hall,  D. J. (1965). \textsl{Isodata: A novel method of data analysis and
			pattern classification}. Stanford Research Institute, Menlo Park.
		\bibitem{}
		Banfield J. and Raftery A. E. (1993). \textsl{Model-based Gaussian and non-Gaussian clustering}. Biometrics. \textbf{49}, 
		803--821.
		\bibitem{}
		Cali\'{n}ski, T. \&  Harabasz, J. (1974). \textsl{A Dendrite Method for Cluster Analysis}. Communications in Statistics -- Theory and Methods. \textbf{3}, 1--27.
		\bibitem{}
		Campello, R. J. G. B., Moulavi, D., Sander, J. (2013). \textsl{Density-Based Clustering Based on Hierarchical Density Estimates}. Proceedings of the 17th Pacific-Asia Conference on Knowledge Discovery in Databases (PAKDD 2013). Lecture Notes in Computer Science. \textbf{7819}, 160--172.
		\bibitem{}
		Celeux, G. and Govaert, G. (1995). \textsl{Gaussian parsimonious clustering models.} Pattern Recognition. \textbf{28}, 781--793.
		\bibitem{}
		Cheng, D., Zhu,  Q., Huang, J., Wu, Q. and Yang, L. (2019).  \textsl{A Novel Cluster Validity Index Based on Local Cores}. IEEE Transactions on Neural Networks and Learning Systems. \textbf{30}, 985--999.	
		\bibitem{}
		Davies,  D. L. and Bouldin,  D. W. (1979). \textsl{A cluster separation measure.} IEEE
		Transactions on Pattern Analysis and Machine Intelligence.
		\textbf{2}, 224--227.
		\bibitem{}
		De, T., Chattopadhyay, T. and Chattopadhyay, A. K. (2014). \textsl{Use of cross-correlation function to study formation mechanism of massive elliptical galaxies}. Publications of the Astronomical Society of Australia, \textbf{31}, Article id: e407, pages 1--8.
		\bibitem{}
		Dempster A. P., Laird N. M., Rubin D. B. (1977). \textsl{Maximum likelihood from incomplete data via the EM algorithm}. 
		Journal of the Royal Statistical Society, Series B. \textbf{39}, 1--38.
		\bibitem{}
		Dunn, J. C. (1974). \textsl{Well-separated clusters and optimal fuzzy partitions.}  Journal of Cybernetics. \textbf{4}, 95--104.
		\bibitem{}
		Efron, B. and Tibshirani, R. (1993). \textsl{An Introduction to the Bootstrap}. Chapman and Hall, New York, London.
		\bibitem{}
		Ester, M., Kriegel, H.-P., Sander, J. \& Xu, X. (1996). \textsl{A density-based algorithm for discovering clusters
			in large spatial databases with noise.} Proceedings of the Second International Conference on
		Knowledge Discovery and Data Mining (KDD-96). AAAI Press, Portland, Oregon, 226--231.
		\bibitem{}
		Everitt, B. S., Landau, S. and Leese,  M. (2001). \textsl{Cluster Analysis.} Arnold, London.
		\bibitem{}
		Flury, B. and Riedwyl, H. (1988).  \textsl{Multivariate Statistics: A practical approach}. Chapman \& Hall, London.
		\bibitem{}
		Frayley, C. and Raftery, A. E. (1998), \textsl{How Many Clusters? Which Clustering
			Method? Answers via Model-Based Cluster Analysis}. The Computer Journal. \textbf{41}, 578--588.
		\bibitem{}
		Fraley C. and Raftery A. E. (1999). \textsl{MCLUST: Software for model-based cluster analysis. Journal of Classification.}, \textbf{16}, 297--306.
		\bibitem{}
		Fraley, C. and Raftery, A. E. (2002). \textsl{Model-based clustering, discriminant analysis and density estimation}. Journal of the American Statistical Association, \textbf{97}, 611--631.
		\bibitem{}
		Fraley, C. and Raftery, A. E. (2003).  \textsl{Enhanced model-based clustering, density estimation, and discriminant analysis 
			software: Mclust.} Journal of Classification. \textbf{20}, 263--286.
		\bibitem{}
		Fraley C. and Raftery A. E. (2007). \textsl{Model-based methods of classification: using the mclust software in 
			chemometrics.} Journal of Statistical Software.\textbf{18},1--13.
		\bibitem{}
		Fraley, C., Raftery, A. E., Murphy, T. B., Scrucca, L. (2012). \textsl{MCLUST version 4 for R: Normal mixture modeling 
			for model-based clustering, classification, and density estimation}. Technical Report. Vol. \textbf{597}, Department of 
		Statistics, University of Washington.
		\bibitem{}
		Handl, J., Knowles, K. \& Kell, D. (2005). \textsl{Computational cluster validation in post-genomic data analysis}. Bioinformatics. \textbf{21}, 3201--3212.
		\bibitem{}
		Hartigan, J. A. (1975). \textsl{Clustering Algorithms}. John Wiley \& Sons, New York, USA.
		\bibitem{}
		Hartigan, J. A. and Wong, M. A. (1979). \textsl{A K-means clustering algorithm}.
		Applied Statistics. \textbf{28}, 100--108.
		\bibitem{}
		Hogg, R. V., Mckean, J. W. and Craig, A. T. (2019). \textsl{Introduction
			to Mathematical Statistics}. Pearson Education, Boston.
		\bibitem{}
		Hope, A. C. A. (1968). \textsl{A simplified Monte Carlo significance test procedure}. Journal of the Royal Statistical Society Series B, \textbf{30}, 582--598.
		\bibitem{}
		Hubert, L. and Arabie, P. (1985). \textsl{Comparing Partitions}, Journal of the Classification, \textbf{2}, 193--218.
		\bibitem{}
		Jain, A. K. , Murty, M. N. and Flynn, P. J. (1999). \textsl{Data clustering: a review}. ACM
		Computing Surveys. \textbf{31}, 264--323.
		\bibitem{}
		Joanes, D. N. and Gill, C. A. (1998). \textsl{Comparing measures of sample skewness and kurtosis}. The Statistician, \textbf{47}, 183--189.
		\bibitem{}
		Johnson, R. A. and Wichern, D. W. (2007). \textsl{Applied Multivariate 
			Statistical Analysis}, Pearson Prentice Hall, New Jersey.
		\bibitem{}
		Jung, S. and Marron,  J. S. (2009). \textsl{PCA consistency in high dimension, low sample size context}. The Annals of Statistics, \textbf{37}, 4104--4130.
		\bibitem{}
		Jure\v{c}kov\'{a}, J. and Kalina, J. (2012). \textsl{Nonparametric multivariate
			rank tests and their unbiasedness.} Bernoulli, \textbf{18}, 229–-251.
		\bibitem{}
		Kass, R. E. and Raftery, A. E. (1995). \textsl{Bayes Factors}. Journal of the American
		Statistical Association. \textbf{90}, 773--795.
		\bibitem{}
		Kaufman, L. and Rousseeuw, P. J. (2005). \textsl{Finding Groups in Data: An Introduction to Cluster Analysis.} John Wiley and Sons, New Jersey.
		\bibitem{}
		Kost, J. T. and McDermott, M. P. (2002). \textsl{Combining dependent p-values.} Statistics \& Probability Letters, \textbf{60}, 183–-190.
		\bibitem{}
		Marozzi, M. (2015). \textsl{Multivariate multidistance tests for high-dimensional low sample size case-control studies.} Statistics in Medicine, \textbf{34}, 1511–-1526.
		\bibitem{}
		Marozzi, M. (2016). \textsl{Multivariate tests based on interpoint distances with application to magnetic resonance imaging.} Statistical
		Methods in Medical Research, \textbf{25}, 2593--2610.
		\bibitem{}
		McLachlan, G. and Peel, D. (2000). \textsl{Finite Mixture Models}. John Wiley and Sons,
		New York.
		\bibitem{}
		Modak, S. (2019). \textsl{Uncovering astrophysical phenomena related to galaxies and other objects through statistical analysis.} Ph.D. Thesis, University of Calcutta, Kolkata, India. URL: http://hdl.handle.net/10603/314773 
		\bibitem{}
		Modak, S. (2021). \textsl{Distinction of groups of gamma-ray bursts in the BATSE catalog through fuzzy clustering}. Astronomy and Computing. \textbf{34}, Article id 100441, Pages 1--7.
		\bibitem{}
		Modak, S. (2022). \textsl{A new nonparametric interpoint distance-based measure for assessment of clustering}. Journal of Statistical Computation and Simulation. \textbf{92}, 1062--1077.
		\bibitem{}
		Modak, S. (2023a). \textsl{Pointwise norm-based clustering of data in arbitrary dimensional space}. Communications in Statistics - Case Studies, Data Analysis and Applications, \textbf{9}, 121–134.
		\bibitem{}
		Modak, S. (2023b). \textsl{Validity index for clustered data in non-negative space}. Calcutta Statistical Association Bulletin, \textbf{75}, 60–71. 
		\bibitem{}
		Modak, S. (2023c). \textsl{A new measure for assessment of clustering based on kernel	density estimation}. Communications in Statistics -- Theory and Methods, \textbf{52}, 5942-5951.
		\bibitem{}
		Modak, S. (2024a). \textsl{A new interpoint distance-based clustering algorithm using kernel density estimation}. Communications in Statistics - Simulation and Computation, \textbf{53}, 5323-5341.
		\bibitem{}
		Modak, S. (2024b). \textsl{Book Review: Finding Groups in Data: An Introduction to Cluster Analysis, Leonard Kaufman \& Peter J. Rousseeuw, 2005}. Journal of Applied Statistics, \textbf{51}, 1618-1620.
		\bibitem{}
		Modak, S. and Bandyopadhyay, U. (2019). \textsl{A new nonparametric test for two sample multivariate location
			problem with application to astronomy.} Journal of Statistical Theory and Applications, \textbf{18}, 136--146.
		\bibitem{}
		Modak, S., Chattopadhyay, A. K. \& Chattopadhyay, T. (2018). \textsl{Clustering of gamma-ray bursts through kernel principal component analysis}. Communications in Statistics -- Simulation and Computation. \textbf{47}, 1088--1102.
		\bibitem{}
		Modak, S., Chattopadhyay, T. \& Chattopadhyay, A. K. (2017). \textsl{Two
			phase formation of massive elliptical galaxies: study through cross--correlation including spatial effect.} Astrophysics and Space Science. \textbf{362}, Article id: 206, Pages 1--10.
		\bibitem{}
		Modak, S., Chattopadhyay, T. \& Chattopadhyay, A. K. (2020). \textsl{Unsupervised classification of eclipsing binary light curves through k-medoids
			clustering}. Journal of Applied Statistics. \textbf{47}, 376--392.
		\bibitem{}
		Modak, S., Chattopadhyay, T. \& Chattopadhyay, A. K. (2022). \textsl{Clustering of eclipsing binary light curves through functional principal component analysis}. Astrophysics and Space Science. \textbf{ 367}, Article id: 19, Pages 1--10.
		\bibitem{}
		Pakhiraa, M. K., Bandyopadhyay, S. and Maulik, U. (2004). \textsl{Validity index for crisp and fuzzy clusters}. Pattern Recognition. \textbf{37}, 487--501.
		\bibitem{}
		Poole, W., Gibbs, D. L., Shmulevich,. I., Bernard, B., Knijnenburg, T. A. (2016). \textsl{Combining dependent P-values with an
			empirical adaptation of Brown’s method.} Bioinformatics, \textbf{32}, i430–-i436.
		\bibitem{}
		Ripley B. D. (1996). \textsl{Pattern recognition and neural networks}. Cambridge University Press, Cambridge.
		\bibitem{}
		Rousseeuw, P. J. (1987). \textsl{Silhouettes: A graphical aid to the interpretation and validation of cluster analysis.} Journal of Computational and Applied Mathematics. \textbf{20}, 53--65. 
		\bibitem{}
		Sch\"{o}lkopf, B. and Smola, A. J. (2002). \textsl{Learning with Kernels: Support Vector Machines, Regularization, Optimization, and Beyond.} MIT Press, Cambridge.
		\bibitem{}
		Schwarz, G. (1978). \textsl{Estimating the Dimension of a Model.} The Annals of Statistics, \textbf{6}, 461--464.
		\bibitem{}
		Scrucca, L., Fop, M., Murphy, T. B. and Raftery, A. E. (2016). \textsl{mclust 5: clustering, classification and
			density estimation using Gaussian finite mixture models}. The R Journal, \textbf{8}, 289--317.
		\bibitem{}
		Silva, L. E. Brito Da, Melton, N. M. and Wunsch, D. C. (2020). \textsl{Incremental Cluster Validity Indices for Online Learning of Hard Partitions: Extensions and Comparative Study}. Institute of Electrical and Electronics Engineers, \textbf{8}, 22025--22047.
		\bibitem{}
		Tarnopolski, M. (2019). \textsl{Analysis of the Duration--Hardness Ratio Plane of Gamma-Ray Bursts Using Skewed
			Distributions.} The Astrophysical Journal. \textbf{870}, 1--9, Article id: 105. 
		\bibitem{}
		Tibshirani, R., Walther, G. \& Hastie, T. (2001). \textsl{Estimating the number of clusters in a data set via the gap statistic}. Journal of the Royal Statistical Society Series B.  \textbf{63}, 411--423.
		\bibitem{}
		T\'{o}th, B. G., R\'{a}cz, I. I. \& Horv\'{a}th, I. (2019). \textsl{Gaussian-mixture-model-based cluster analysis of gamma-ray bursts in
			the BATSE catalog}. Monthly Notices of the Royal Astronomical Society. \textbf{486}, 4823--4828.
		\bibitem{}
		Vale, D. C. and Maurelli V. A. (1983). \textsl{Simulating multivariate nonnormal distributions}. Psychometrika, \textbf{48}, 465--471.
		\bibitem{}
		Yata, K. and Aoshima, M. (2010). \textsl{Effective PCA for high-dimension, low-sample-size data with singular
			value decomposition of cross data matrix}, Journal of Multivariate Analysis, \textbf{101},  2060--2077.
	\end{thebibliography}
\end{document}